\documentclass[11pt]{article}

\usepackage{cite}
\usepackage{amsmath,amssymb,amsfonts}
\usepackage{graphicx}
\usepackage{placeins}
\usepackage{subcaption}
\graphicspath{{figures/}}
\usepackage{textcomp}
\usepackage{multirow}
\usepackage{csquotes}
\usepackage{xcolor}
\usepackage{float}
\usepackage{url}
\title{On the Representational Limits of Quantum-Inspired 1024-D Document Embeddings: An Experimental Evaluation Framework}

\author{
Dario Maio\\
Department of Computer Science and Engineering\\
University of Bologna, Italy\\
\texttt{dario.maio@unibo.it}
}

\date{}

\begin{document}

\maketitle

\begin{abstract}
Text embeddings are central to modern information retrieval and Retrieval-Augmented Generation (RAG). While dense models derived from Large Language Models (LLMs) dominate current practice, recent work has also explored quantum-inspired alternatives, motivated by their geometric richness and potential efficiency. This paper presents an experimental framework for constructing \textbf{quantum-inspired 1024-dimensional document embeddings}, based on overlapping text windows and designed for retrieval tasks.

The main contributions of this work can be summarized as follows. We first introduce an experimental pipeline that combines overlap and multiscale processing with optional quantum-inspired components (such as EigAngle and quasi-kernel sampling), together with linear or MLP-based distillation and JSON-based fingerprinting for reproducibility. We then present a set of diagnostic tools for hybrid retrieval, including both static and dynamic interpolation between BM25 and embedding-based scores, candidate union strategies, and a conceptual $\alpha$-oracle that provides an upper bound for score-level fusion. In this setting, $\alpha=0$ corresponds to pure BM25, while $\alpha=1$ relies entirely on embeddings. Finally, we report empirical results on controlled corpora of Italian and English documents across technical, narrative, and legal domains, using synthetic queries, and complement these with sentence-level evaluations comparing QEMB and teacher embedding similarity.

Across these settings, BM25 is generally a very strong baseline, teacher embeddings are often more stable than QEMB, and standalone quantum-inspired embeddings remain insufficient for robust retrieval. Distillation shows mixed effects: in some settings it improves standalone QEMB retrieval, while in others it provides limited gains or weakens hybrid effectiveness. Hybrid fusion can sometimes recover competitive performance, especially when quantum-inspired signals complement lexical matching, but these benefits are not uniform across datasets or query formulations. At the geometric level, alignment between quantum-inspired and teacher embedding spaces remains weak or unstable in several settings, suggesting structural limitations in the encoding that are only partially mitigated by distillation.

Beyond the empirical results, this study introduces an experimental evaluation framework and a set of diagnostic tools to analyze how embedding-space geometry, distance compression, and ranking instability influence retrieval behavior. These analyses help clarify both the limitations and the potential supporting role of quantum-inspired embeddings in document retrieval.
\end{abstract}


\section{Introduction}\label{sec:intro}
In recent years, Large Language Models (LLMs) have become central to modern artificial intelligence (AI), with a growing impact on tasks such as question answering, summarization, code generation, and Retrieval-Augmented Generation (RAG).

Surveys such as Minaee \textit{et al.}~\cite{minaee2024largelanguagemodelssurvey} highlight how improvements in model architecture, scaling strategies, and training data have strengthened the role of LLMs as representation models, where text embeddings are a key component for downstream reasoning and retrieval.
In particular, RAG has become a widely adopted approach for integrating external knowledge into LLM-based systems.

Recent surveys provide a comprehensive overview of RAG techniques and their integration with large models \cite{gao2024retrievalaugmentedgenerationlargelanguage,fan2024surveyragmeetingllms}, highlighting how retrieval quality and document embeddings critically affect downstream performance.

Despite their effectiveness, LLM embeddings also have limitations: they are computationally expensive, often language-specific, and difficult to interpret. 
These shortcomings motivate the exploration of alternative paradigms, including quantum computing and quantum-inspired models. 
Such approaches draw on the geometric richness of Hilbert spaces, where superposition and interference enable the representation of multiple semantic hypotheses within a single state. 
In contrast to classical embeddings, which map text to a single point in vector space, quantum-inspired representations can be interpreted as distributions over latent semantic configurations, potentially capturing ambiguity and contextual overlap in a more expressive way.

However, whether these geometric properties translate into meaningful similarity structure for retrieval remains an open question, which this work investigates empirically.

This paper investigates whether a \textbf{quantum-inspired pipeline producing fixed 1024-dimensional embeddings} can support document retrieval, and to what extent such representations exhibit structural limitations when compared with strong lexical and dense baselines. In this work, QEMB encodes each text segment through a window-based decomposition, where each window is mapped into a fixed-size feature vector via angle projection followed by a quantum-inspired transformation. The resulting window-level features are aggregated into a single 1024-dimensional embedding.

In the current experimental configuration, each sub-chunk is divided into a fixed number of windows producing feature vectors that together form a 1024-dimensional representation, further refined through overlap and dense multi-scale fusion before final L2 normalization.

The use of a 1024-dimensional embedding space is a deliberate design choice, driven by the compositional structure of the QEMB pipeline and aligned with the dimensionality of the teacher model, while remaining consistent with common practice in modern embedding systems.

High-dimensional representation spaces are known to exhibit non-intuitive geometric effects, often discussed under the curse of dimensionality \cite{aggarwal2001curse}.

Unlike prior work that compresses or augments pretrained embeddings, we analyze embeddings constructed directly from text windows, mapped through deterministic feature extractions and optional quantum-inspired transformations, including semantic projections (EigAngle) and quasi-kernel sampling mechanisms.

To this end, we design controlled experimental setups based on corpora of Italian and English documents spanning technical, narrative, and legal domains, paired with synthetic queries. A representative subset of results is reported, while a larger set of experiments has been conducted to validate the observed trends. These datasets highlight the challenge of obtaining high-quality embeddings across different languages and domains. Although modest in scale, they are designed to enable controlled and interpretable diagnostic analysis. Selected experimental resources are available upon reasonable request, subject to data-sharing constraints. The goal of this work is not full reproducibility in the sense of artifact release, but methodological transparency and controlled diagnostic evaluation. The experimental setup is fully specified in terms of pipeline configuration, similarity regimes, and evaluation protocol, allowing consistent analysis of embedding behavior across different conditions. Our evaluation covers BM25 \cite{robertson2009bm25}, FAISS-indexed teacher embeddings \cite{johnson2019faiss}, and our quantum-inspired embeddings, both raw and distilled. We compare multiple fusion strategies, including score-level interpolation controlled by $\alpha$ (both static and dynamically gated), rank-based fusion using Reciprocal Rank Fusion (RRF), candidate-union retrieval, and cross-encoder re-ranking with guard rails.
We also provide detailed per-query diagnostics, identifying failure modes such as candidate loss, Doc$\to$Sub degradation, and cross-encoder regressions. In addition to retrieval-level evaluation, we conduct controlled sentence-level experiments to assess whether quantum-inspired embeddings preserve semantic similarity. These pairwise evaluations provide a direct probe of the geometric properties of the embedding space, independently of retrieval dynamics.

Rather than introducing a new state-of-the-art method, this work adopts a diagnostic perspective. 
We first investigate whether quantum-inspired embeddings can faithfully represent semantic similarity at a local, pairwise level. 
We then analyze how these geometric properties propagate to document retrieval performance.

This two-step approach allows us to disentangle representational limitations from retrieval dynamics and to better understand the role of quantum-inspired embeddings within hybrid search pipelines.

\section{Contributions}\label{sec:contrib}
The main contributions of this work are:

\begin{itemize}
  \item An experimental framework for 1024-dimensional quantum-inspired document embeddings, based on sub-chunk and window-level encoding with overlap, multi-scale aggregation, and optional semantic projections.

  \item A unified evaluation framework for hybrid retrieval, combining BM25, dense embeddings, and score-level fusion, with consistent metrics across multiple corpora and controlled experimental settings.

 \item An empirical analysis of quantum-inspired embeddings across technical, narrative, and legal corpora, showing that:
 (i) standalone QEMB embeddings provide weak and unstable ranking signals;
 (ii) linear and Multi-Layer Perceptron (MLP) distillation have mixed effects, improving standalone QEMB retrieval in some settings while not consistently improving, and sometimes degrading, hybrid performance;
 (iii) hybrid retrieval can recover competitive performance in some settings when combining lexical and embedding-based signals, but does not consistently outperform strong lexical baselines.

  \item A set of diagnostic insights linking geometric properties of embedding spaces to retrieval behavior, highlighting the role of distance compression and ranking instability in limiting standalone embedding performance.

 \item Controlled experimental setups spanning multiple domains and languages enable consistent diagnostic analysis of embedding behavior under different semantic regimes.
 
\item Analysis on selected corpora of Italian and English documents spanning technical, narrative, and legal domains with synthetic queries, highlighting challenges in multilingual and multi-domain retrieval.
\end{itemize}

In addition, we provide a detailed analysis of score-level fusion mechanisms. We study the score-level fusion between BM25 and embeddings, controlled by a weight parameter $\alpha$. 
Specifically, $\alpha=0$ corresponds to pure BM25, $\alpha=1$ to embeddings only, and intermediate values blend the two signals. 
We evaluate score-level fusion controlled by $\alpha$, including both static settings across $\{0,0.1,0.2,0.3,0.5,0.8,1\}$ and a dynamic $\alpha$ gated by BM25 confidence, together with rank-based fusion using Reciprocal Rank Fusion (RRF). The optimal value of $\alpha$ is not fixed, and varies across datasets and query distributions, reflecting the relative contribution of lexical and embedding-based signals.
We also introduce an \emph{$\alpha$--oracle}, defined as the upper bound achievable by score interpolation before applying cross-encoder (CE) re-ranking. In this work, it is used as a conceptual diagnostic reference to interpret the potential headroom of score-level fusion, rather than as an explicitly optimized quantity.

The remainder of this paper is organized as follows.
Section~\ref{sec:related} reviews related work.
Section~\ref{sec:representation} discusses the representational capacity of quantum-inspired embedding spaces, including direct quantum-inspired embeddings (Section~\ref{sec:direct_qemb}), quantum transformations over classical embeddings (Section~\ref{sec:quantum_transformations}), and kernel-based approaches (Section~\ref{sec:kernel_methods}).
Section~\ref{sec:pipeline} presents the QEMB framework, the experimental setup, and the retrieval results on technical, narrative, and legal corpora.
Finally, Section~\ref{sec:limitations} discusses limitations, followed by Section~\ref{sec:conclusions}, which presents conclusions and future research directions.

\section{Related Work}\label{sec:related}

\noindent\textbf{Quantum-inspired IR and representational formalisms}
A long line of research has explored quantum-theoretic formalisms for information retrieval (IR)-from meaning representations and interference/superposition effects to density-matrix modeling.
Aerts et al.\ proposed a meaning-focused, quantum-inspired view of IR where documents and queries are states in a Hilbert space and composition yields interference phenomena \cite{aerts2013meaning}.
Earlier, Huertas-Rosero \textit{et al.}\ framed lexical operations as ``measurements'' that partially erase information, hinting at intrinsic trade-offs in representation and retrieval \cite{huertas2008erasing}.
A comprehensive survey by Uprety \textit{et al.}\ covers quantum-like probabilistic models, projectors, and interference-based ranking \cite{uprety2020survey}.
Recent work has explored the integration of quantum-inspired retrieval mechanisms in hybrid systems, although this direction remains relatively underexplored and still lacks methodological consolidation \cite{singh2025quantumragandpungpt2}.

\noindent\textbf{Quantum-inspired embeddings and similarity metrics.}
Quantum-inspired complex-valued word embeddings have been explored for modeling lexical semantic representations \cite{li2018complex}.
Kankeu \textit{et al.}\ proposed a quantum-inspired projection head that compresses BERT embeddings \cite{devlin2019bert} via simulated circuits and evaluates similarity through fidelity; on the TREC Deep Learning benchmark \cite{craswell2020trec}, they report competitive performance with far fewer parameters than classical heads \cite{kankeu2025qiepsm}.
A foundational contribution in this direction is the work of Lloyd \textit{et al.}, who proposed training quantum encoding circuits to maximize class separation in Hilbert space, establishing a theoretical basis for quantum metric learning \cite{lloyd2020quantum}.

\noindent\textbf{Quantum-inspired models for semantic matching, fusion, and density-matrix-based representations} have also been explored \cite{zhang2025density_matching,gao2025qsim,duan2024qfm}.
\noindent\textbf{Quantum-inspired word embeddings.}
Li explored complex-valued word embeddings with quantum motivations \cite{li2018complex}.

\noindent\textbf{Hybrid retrieval and rank/score fusion.}
Hybrid lexical--semantic retrieval is now standard practice. Reciprocal Rank Fusion provides a robust combination strategy \cite{cormack2009rrf}. Quantum-like interference effects have also been explored in decision fusion for ranking multimodal documents \cite{gkoumas2018fusion}.

\noindent\textbf{Distillation and embedding compression.}
Teacher–student training, projection heads, and autoencoding-based embedding refinement 
have been widely explored\cite{reimers2019sentencebert,kaneko2020autoencoding,zhao2022hpd}.
In the quantum domain, Shi \textit{et al.}\ explored classical-to-quantum knowledge distillation, using a classical teacher to guide quantum circuit optimization, improving robustness on real quantum hardware \cite{shi2023nqe}.
Taken together, prior work has mainly explored quantum-inspired methods as alternative representational or similarity mechanisms, often with the goal of improving effectiveness, compression, or modeling flexibility. In contrast, the present work adopts a diagnostic perspective: rather than proposing a new competitive retrieval model, it investigates the structural limits of high-dimensional quantum-inspired embeddings and analyzes how their geometric properties affect retrieval behavior across controlled settings.

\section{Semantic Representational Capacity of Hilbert-like Spaces}
\label{sec:representation}

This section investigates whether quantum-inspired embedding spaces can effectively represent semantic similarity relevant to retrieval.

We distinguish three paradigms, which will be treated separately in the following subsections:

\begin{itemize}
\item \textbf{Direct quantum-inspired embeddings}, constructed from text through deterministic or circuit-based encodings;
\item \textbf{Quantum transformations over classical embeddings}, where a semantically meaningful space is mapped into a Hilbert space;
\item \textbf{Kernel-based approaches}, which define similarity implicitly but do not produce indexable embeddings.
\end{itemize}

While all three paradigms are theoretically grounded, only embedding-based approaches are directly compatible with scalable retrieval systems. This motivates an empirical investigation of representational quality before considering retrieval performance.

\subsection{Direct Quantum-Inspired Embeddings}
\label{sec:direct_qemb}

We evaluate the geometric properties of embedding spaces using controlled sentence pairs, focusing on a representative subset of configurations selected from a broader set of experiments.

\subsubsection{Pairwise Similarity Evaluation Setup}

We evaluate the geometric properties of embedding spaces using a controlled set of sentence pairs.

\begin{itemize}
\item \textbf{Dataset:} A small, manually curated set of 10 sentence pairs in English, designed as a controlled diagnostic probe. The pairs are selected to cover three similarity regimes (\textit{sim}, \textit{neutral}, \textit{dissim}), including extended paraphrases, partial semantic overlap, and semantically unrelated sentences.
\item \textbf{Teacher model:} \texttt{intfloat/multilingual-e5-large}~\cite{wang2024multilingual}, a multilingual embedding model with 1024-dimensional output vectors, 24 layers, and a maximum input length of 512 tokens (longer inputs are truncated).

\item \textbf{Embedding spaces:}
  \begin{itemize}
    \item \textbf{QEMB (baseline configuration)}: the baseline QEMB encoding used in the current experimental setup, based on fixed window decomposition and deterministic feature extraction
    \item \textbf{QEMB (amplitude, experimental)}: a variant using amplitude-based encoding, evaluated as an exploratory alternative
    \item \textbf{Distilled embeddings}: not included in this experiment
  \end{itemize}
\item \textbf{Ground truth:} LLM-based similarity score
\end{itemize}

\paragraph{LLM-based similarity proxy.}
Similarity scores used as reference in the pairwise evaluation are generated using a large language model (LLM), 
specifically \texttt{GPT-5.2} (OpenAI), configured in a controlled generation setting.\footnote{GPT-5.2 refers to a proprietary large language model developed by OpenAI. Public information about the model family is available at \url{https://platform.openai.com/docs}.}

The LLM is prompted to produce sentence pairs together with a scalar similarity score explicitly instructed to approximate cosine similarity between embeddings, rather than deep semantic relatedness. The prompt enforces:
(i) reliance on lexical overlap and surface similarity,
(ii) avoidance of narrative or inferential reasoning,
(iii) predefined similarity ranges corresponding to different regimes (high, medium, low, dissimilar).

Sentence pairs and scores are generated in a structured TSV format. The generation process includes multiple layers of control:
\begin{itemize}
\item predefined distribution across similarity regimes,
\item token-length constraints (20–50 tokens per sentence),
\item temperature-controlled sampling,
\item strict parsing and validation of output format,
\item post-generation filtering to enforce token-length constraints using a tokenizer-based check.
\end{itemize}

\textbf{Important limitation.}
The LLM-based score is not intended as a semantic ground truth in the human sense. Instead, it serves as a controlled proxy designed to approximate cosine-like similarity under constrained generation conditions. In particular, the prompting procedure explicitly emphasizes lexical overlap and surface-level similarity, rather than deeper semantic reasoning.

This design differs from standard STS evaluation: the goal is not to benchmark semantic understanding, but to probe whether the embedding space preserves relative similarity structure under controlled conditions. Accordingly, the LLM-based score should be interpreted as a diagnostic reference signal for geometric alignment, not as an absolute measure of meaning. The generation procedure is specified at the methodological level, including similarity regimes, output constraints, and filtering criteria, enabling comparable diagnostic settings to be reconstructed.

\subsubsection{Metrics}

We evaluate the alignment between embedding-based similarity and reference similarity scores using a set of complementary metrics.

\textbf{Correlation measures.}
We report both Pearson and Spearman correlation coefficients between embedding cosine similarity and LLM-based similarity scores. 
Pearson correlation captures linear alignment, while Spearman correlation measures rank consistency, providing a more robust indicator of relative ordering.

\textbf{Absolute error.}
We compute the mean absolute error (MAE) between embedding similarity and LLM scores, quantifying the deviation from the reference signal.

\textbf{Similarity statistics.}
To assess discriminative behavior, we report aggregate statistics of similarity scores (e.g., mean similarity), which help identify effects such as score compression and lack of separation between semantic classes.

All metrics are computed consistently across all evaluated representations, including direct QEMB embeddings and exploratory ZZFeatureMap-based probes.

\subsubsection{Pairwise Similarity Results}

The results highlight a fundamental geometric limitation of the QEMB encoding.

\paragraph{Quantitative results.}
Table~\ref{tab:pairwise_results} reports correlation and error metrics across the evaluated representations.

\begin{table}[t]
\centering
\caption{Pairwise similarity evaluation across embedding spaces.}
\label{tab:pairwise_results}
\begin{tabular}{lcccc}
\hline
\textbf{Space} & \textbf{Pearson} & \textbf{Spearman} & \textbf{MAE} & \textbf{Mean sim.} \\
\hline
Teacher (E5)              & 0.97 & 0.95 & 0.44 & 0.88 \\
QEMB (baseline)           & -0.25 & -0.44 & 0.54 & 0.87 \\
QEMB (amplitude, exp.)    & 0.58 & 0.65 & 0.44 & 0.84 \\
\hline
\end{tabular}
\end{table}

\noindent\textit{The results indicate a clear degradation in similarity ordering for the baseline QEMB configuration, while the amplitude-based variant partially improves correlation but remains far from the teacher baseline.}

Teacher embeddings achieve near-perfect alignment with the reference similarity scores, confirming the validity of the evaluation setup.

It is worth noting that the MAE of the Teacher embedding (0.44) may appear relatively high despite the near-perfect correlation. This is explained by a difference in scale between the two signals: cosine similarity values tend to concentrate in the high-similarity region, while LLM-based scores are encouraged to span a wider range. As a result, even a rank-consistent embedding can exhibit non-negligible absolute deviations. This highlights that MAE should be interpreted jointly with correlation measures rather than as a standalone indicator of alignment quality.

In contrast, the QEMB \emph{baseline} configuration exhibits a complete breakdown of alignment, with negative correlation and high error. This indicates not merely a loss of precision, but a pathological inversion of similarity structure, in which semantically related and unrelated pairs are no longer ordered coherently.

The amplitude-based variant partially mitigates this issue, achieving moderate correlation and lower error, but still remaining far from the teacher baseline.

\paragraph{ZZ-based transformations.}
In addition to the main embedding configurations, we evaluate ZZ-based transformations, whose results are reported in Table~\ref{tab:zz_pairwise_results}.

\begin{table}[t]
\centering
\caption{Pairwise similarity evaluation for ZZ-based transformations.}
\label{tab:zz_pairwise_results}
\begin{tabular}{lcccc}
\hline
\textbf{Space} & \textbf{Pearson} & \textbf{Spearman} & \textbf{MAE} & \textbf{Mean sim.} \\
\hline
ZZ (raw)            & 0.04 & -0.36 & 0.73 & 0.00 \\
ZZ (teacher)        & 0.46 & 0.22  & 0.52 & 0.97 \\
ZZ (window raw)     & 0.34 & 0.55  & 0.45 & 0.80 \\
\hline
\end{tabular}
\end{table}

\noindent\textit{ZZ-based transformations exhibit highly unstable geometric behavior: the raw variant collapses toward near-zero similarity, the teacher-based transformation produces strong score compression near 1.0, while the window-based variant yields more balanced but still suboptimal correlation.}

\begin{figure}[t]
\centering
\includegraphics[width=0.75\linewidth]{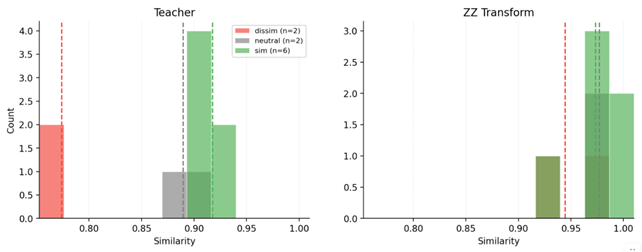}
\caption{Distribution of similarity scores for ZZ-based transformations compared with teacher embeddings. 
The distribution shows strong concentration in the high-similarity region, with limited separation between semantic classes.}
\label{fig:zz_teacher}
\end{figure}

\begin{figure}[t]
\centering
\includegraphics[width=0.75\linewidth]{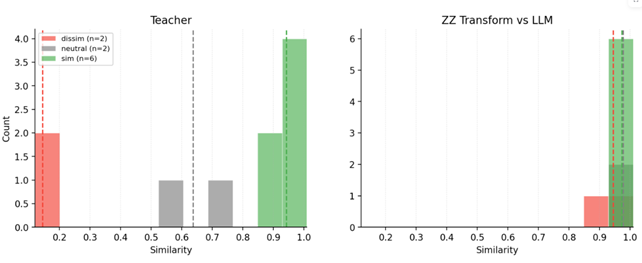}
\caption{Distribution of similarity scores for ZZ-based transformations compared with LLM-based similarity scores. 
The compression of similarity values reduces discriminative power across semantic classes.}
\label{fig:zz_llm}
\end{figure}

This behavior is also evident in the distribution of similarity scores, as shown in Figures~\ref{fig:zz_teacher} and~\ref{fig:zz_llm}.

\paragraph{Similarity structure.}
All QEMB-based representations tend to produce uniformly high similarity scores, independently of the semantic relation between sentence pairs. 
Even dissimilar pairs are assigned high similarity values, leading to a strong compression of distances in the embedding space.

\paragraph{Interpretation.}
These results indicate that the main limitation of the QEMB encoding lies in its inability to preserve relative similarity structure. 
Rather than producing a meaningful ranking of sentence pairs, the embedding space collapses toward uniformly high similarity values, resulting in poor discriminative power.

The amplitude-based variant introduces slightly more variability, but does not recover the semantic structure observed in teacher embeddings.

\paragraph{Comparative analysis.}
Across all configurations, QEMB embeddings consistently exhibit distance compression effects, particularly for dissimilar pairs, which are assigned unexpectedly high similarity values.

This behavior is visually confirmed in Figures~\ref{fig:zz_teacher} and~\ref{fig:zz_llm}, where ZZ-based transformations produce a strong concentration of similarity scores in the high-value region, leading to significant overlap between semantic classes.

While ZZ transformations slightly alter the distribution of similarity values, they do not recover a stable or semantically meaningful geometry, and fail to restore class separability.

All metrics are computed consistently across all evaluated representations.

\subsection{Quantum Transformations over Classical Embeddings}
\label{sec:quantum_transformations}

In this paradigm, a semantically meaningful classical embedding space is explicitly transformed into a new representation through quantum-inspired mappings. Unlike direct quantum-inspired embeddings, the starting point is a vector space that already encodes semantic similarity (e.g., teacher embeddings), and the transformation produces a new set of vectors in a Hilbert-like space.

From a theoretical perspective, semantic structure is not constructed from scratch, but inherited from the underlying classical model. As a consequence, the transformation is expected to preserve, at least partially, the relative similarity ordering of the original space. However, this behavior depends critically on the stability of the mapping: strong non-linearities or distance distortions introduced by the quantum-inspired transformation may degrade or destroy the inherited structure.

A further implication is that the resulting representations are not standalone. The embedding function remains tied to the underlying teacher model and, in some configurations, to the data used to define the transformation. This limits generalization across datasets and tasks, in contrast to standard embedding models, which learn transferable representations that can be applied across different corpora and tasks. 

Consequently, while theoretically elegant, quantum kernel methods remain computationally prohibitive for the millisecond-latency requirements of modern RAG pipelines.

Crucially, this paradigm still produces explicit vector representations, which can be indexed and integrated into standard retrieval pipelines. This distinguishes it from kernel-based approaches, where similarity is defined implicitly and indexing becomes more challenging.
From an experimental perspective, this setting is partially covered in Section~\ref{sec:direct_qemb} through the \texttt{zz\_teacher} configuration, where a ZZFeatureMap is applied to teacher embeddings. The results reported in Table~\ref{tab:zz_pairwise_results}, obtained on the dataset described in Section~\ref{sec:direct_qemb}, show that, even when starting from a semantically meaningful space, the resulting geometry remains unstable.

In particular, score compression and weak correlation with the reference signal indicate that the transformation does not reliably preserve similarity structure. Consistent behavior is observed in additional experiments on a small set of legal-domain sentence pairs, suggesting that this instability is not limited to the narrative setting.

These findings suggest that the limitations of this paradigm are not only empirical, but reflect a structural tension between the geometry induced by quantum-inspired mappings and the metric properties required for stable similarity preservation.

\subsection{Kernel-Based Approaches}
\label{sec:kernel_methods}

Kernel-based approaches define similarity implicitly through inner products in high-dimensional vector spaces and, in quantum settings, Hilbert spaces, without necessarily constructing explicit embeddings. This formulation naturally arises in quantum machine learning frameworks \cite{schuld2021qml}.

In contrast to transformation-based methods, similarity is computed via a kernel function, which may operate either directly on input representations or on implicitly mapped feature spaces. While explicit feature maps can sometimes be derived, they are often high-dimensional, data-dependent, or computationally expensive to materialize.

From a practical perspective, indexing is not impossible, but becomes significantly more challenging. 
In particular, the lack of a stable, low-dimensional, and explicitly constructed embedding space limits the applicability of standard approximate nearest neighbor (ANN) methods and complicates large-scale retrieval.

Another key limitation is that many kernel constructions are strongly corpus-dependent. 
In practice, the resulting similarity function is tied to the specific dataset used to define or approximate the kernel, rather than representing a reusable mapping from text to a stable embedding space.

As a consequence, kernel methods do not behave like standard embedding models, which learn transferable representations that can be applied across different corpora and tasks.
As a result, these methods do not naturally generalize as reusable embedding functions, in contrast to standard neural embedders trained to produce transferable representations across datasets and domains.

Moreover, kernel evaluation in such settings typically requires pairwise computations between queries and corpus elements, leading to poor scalability as corpus size increases.
This limitation applies in particular to kernel constructions based on explicit feature maps or circuit-based transformations, where the similarity cannot be directly indexed using standard approximate nearest neighbor methods.

In addition to these constructions, alternative quantum-inspired formulations have been proposed in which interference-like effects are introduced directly at the level of the similarity function, for instance by augmenting inner products with phase-based components, without modifying the underlying embedding space.

In the current experimental configuration, the quantum kernel is implemented by mapping input embeddings into a low-dimensional space compatible with the circuit width. Specifically, high-dimensional embedding vectors are projected via Principal Component Analysis (PCA) \cite{jolliffe2002pca} onto a number of components equal to the number of available qubits. 
This construction fundamentally differs from similarity-level quantum-inspired formulations, as the kernel is computed through an explicit feature mapping and circuit-based transformation, rather than as a modification of the inner product in the original embedding space.

The resulting vectors are encoded using angle encoding and processed by a quantum circuit simulated on the Aer backend. Kernel values are estimated as the inner product between the corresponding quantum states.

This construction introduces an information bottleneck: the dimensionality reduction imposed by PCA, together with the limited number of qubits, constrains the expressiveness of the kernel and may hinder the preservation of semantic structure. These design choices directly inform the experimental results reported below.

For these reasons, kernel-based methods are best interpreted as analytical or experimental tools for studying similarity functions, rather than as primary components of large-scale retrieval pipelines.

Improvements observed in hybrid systems incorporating quantum-inspired modifications at the level of the similarity function, rather than through explicit feature maps or Hilbert-space transformations, may partly arise from non-linear score transformations and ensemble effects, rather than from a genuine increase in representational expressiveness.

\paragraph{Experimental results.}
Table~\ref{tab:qk_results} reports the correlation between quantum kernel similarity, teacher similarity, and LLM-based scores.

\begin{table}[t]
\centering
\caption{Quantum kernel similarity evaluation.}
\label{tab:qk_results}
\begin{tabular}{lcc}
\hline
\textbf{Comparison} & \textbf{Pearson} & \textbf{Spearman} \\
\hline
Teacher vs LLM      & 0.97 & 0.95 \\
Quantum vs Teacher  & 0.25 & 0.18 \\
Quantum vs LLM      & 0.16 & 0.24 \\
\hline
\end{tabular}
\end{table}

\begin{figure}[!t]
\centering

\includegraphics[width=0.85\linewidth]{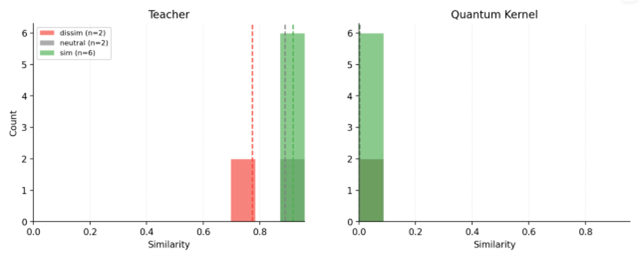}

\vspace{4mm}

\includegraphics[width=0.85\linewidth]{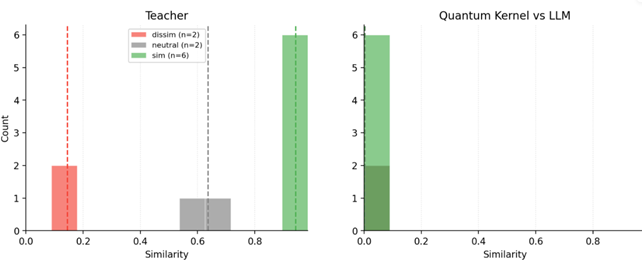}

\caption{Distribution of similarity scores for the quantum kernel compared with teacher embeddings (top) and LLM-based similarity scores (bottom). 
The similarity distribution collapses toward near-zero values, significantly reducing separability across semantic classes.}
\label{fig:qk_combined}
\end{figure}

These results indicate that the quantum kernel struggles to preserve the semantic structure of the original embedding space. 
While some weak correlation with teacher and LLM similarity is retained, the overall similarity distribution collapses toward near-zero values, significantly reducing discriminative power.

This behavior is clearly visible in Figure~\ref{fig:qk_combined}, where the similarity distribution concentrates in a narrow region with minimal separation between semantic classes.

Additional qualitative evidence is provided by experiments conducted on a separate set of legal-domain sentence pairs (in Italian), distinct from the dataset described in Section~\ref{sec:direct_qemb}. The results are consistent with the behavior observed in the main evaluation and further illustrate the instability of the quantum kernel.
\begin{figure}[t]
\centering
\includegraphics[width=0.85\linewidth]{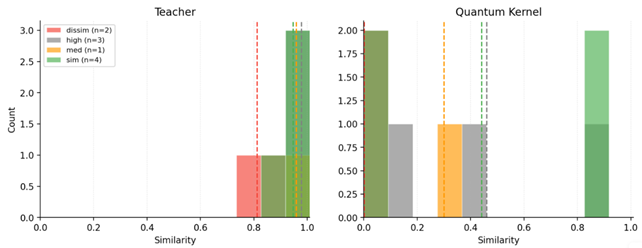}
\caption{Distribution of similarity scores for the quantum kernel compared with teacher embeddings on a small set of legal-domain sentence pairs. Unlike the teacher model, which preserves a meaningful separation between semantic classes, the quantum kernel collapses most similarities toward zero, leading to near-orthogonal representations and loss of discriminative power.}
\label{fig:qk_legal}
\end{figure}

This behavior confirms that the degeneracy observed in the quantum kernel is not specific to the narrative dataset, but persists across domains, including legal text.

\FloatBarrier
\paragraph{Summary of Section 4.}
Overall, the three paradigms highlight different trade-offs between representational expressiveness, computational cost, and generalization. 
Direct quantum-inspired embeddings aim to construct representations from raw text but suffer from limited semantic alignment; transformation-based approaches operate on semantically meaningful spaces but remain dependent on the underlying embedding model; kernel-based methods offer a flexible similarity framework but lack scalability and transferable representations.
From a geometric perspective, kernel-based approaches arguably provide the most coherent similarity formulation, as they operate directly in a Hilbert space without requiring the construction of a compressed explicit embedding. However, this advantage comes at a significant practical cost: the lack of indexable representations, strong dependence on corpus-specific transformations, and high computational complexity make them unsuitable for scalable retrieval pipelines.

These distinctions motivate the experimental analysis that follows, focusing on how such geometric properties translate into retrieval behavior within hybrid pipelines.

\section{QEMB Framework and Embedding Pipeline}
\label{sec:pipeline}

The QEMB framework implements an experimental pipeline for constructing 1024-dimensional embeddings.

\subsection{From Geometry to Retrieval}
\label{sec:geometry_to_retrieval}

The results of Section~\ref{sec:direct_qemb} show that quantum-inspired representations exhibit significant geometric distortions, including distance compression, weak alignment with semantic similarity, and unstable similarity distributions.

These effects have direct implications for retrieval. In particular:

\begin{itemize}
\item \textbf{Distance compression} reduces score contrast, making it difficult to discriminate between relevant and non-relevant documents;
\item \textbf{Weak alignment with semantic similarity} leads to incorrect ranking of candidate documents;
\item \textbf{Similarity collapse or saturation} (as observed in ZZ and kernel-based transformations) further degrades ranking quality by reducing effective dynamic range;
\item \textbf{Local similarity errors} propagate through multi-stage retrieval pipelines, affecting both candidate selection and re-ranking.
\end{itemize}

As a consequence, retrieval performance cannot be understood independently of the underlying embedding geometry. This motivates a systematic evaluation of retrieval behavior, aimed at quantifying how these geometric limitations impact end-to-end retrieval effectiveness.

\subsection{Pipeline Overview}

The QEMB encoder processes each input text segment (sub-chunk) by dividing it into a variable number of windows.
In this work, the term "quantum-inspired" refers to classical simulations of quantum feature mappings, in which text-derived parameters are encoded through transformations inspired by quantum state preparation and evolution. Here, the "quantum-inspired" aspect is primarily reflected in the geometry of the induced feature space, rather than in any direct quantum computational advantage. This includes both angle-based encodings, in which features parameterize rotation operations, and amplitude-based representations, where normalized feature vectors are interpreted as state amplitudes. These representations are processed through circuit-inspired feature extraction, that mimics structural properties of quantum circuits (e.g., parameterized rotations and entanglement patterns), while remaining fully implementable on classical backends.
For each window, token-level statistics are projected into angle parameters and then processed by a quantum-inspired circuit, implemented either via the Aer simulator or a Torch-based surrogate. The resulting window-level feature vectors are then aggregated and resampled to a fixed number of windows before concatenation, producing a 1024-dimensional embedding.
Each sub-chunk embedding is obtained as the concatenation of fixed-size window-level feature vectors. 
Formally, given $K$ window-level feature vectors and $F$ features per window, the embedding is constructed from a resampled set of $W=16$ windows:

\paragraph{Window aggregation and fixed-dimensional embedding.}
Given a set of window-level feature vectors $\{\mathbf{f}_i\}_{i=1}^{K}$, where $K$ depends on the sub-chunk length and windowing strategy, the final embedding is constructed through a fixed resampling and concatenation procedure.

Since the number of windows per sub-chunk may vary, a deterministic resampling step maps the sequence of window features to a fixed number $W=16$ of slots. This is implemented via a deterministic aggregation strategy over contiguous windows, which may include pooling, interpolation, or projection-based resampling.
\[
\tilde{\mathbf{f}}_j = \mathrm{Aggregate}\big(\{\mathbf{f}_i\}_{i \in \mathcal{I}_j}\big), \quad j = 1, \dots, 16,
\]
where $\{\mathcal{I}_j\}$ defines a partition of the original window sequence.

Although various aggregation strategies can be used, they all share a common constraint: the number of output windows is fixed.

The final embedding is obtained by concatenation:
\[
\mathbf{e} = [\tilde{\mathbf{f}}_1 \; \Vert \; \tilde{\mathbf{f}}_2 \; \Vert \; \dots \; \Vert \; \tilde{\mathbf{f}}_{16}] \in \mathbb{R}^{1024},
\]
with each $\tilde{\mathbf{f}}_j \in \mathbb{R}^{64}$.

Finally, the embedding is L2-normalized:
\[
\mathbf{e} \leftarrow \frac{\mathbf{e}}{\|\mathbf{e}\|_2}.
\]

This procedure ensures a fixed embedding dimensionality (1024), independently of the sub-chunk length and the number of initial windows.

This encoding is deterministic with respect to the input text and parameter configuration, and does not require gradient-based training. The proposed framework is not centered on end-to-end learning, but rather on analyzing the representational properties of a deterministic encoding, with learning introduced only through optional distillation.
Angle parameters used by the encoder may originate from three alternative sources: lexical projections, lightweight semantic projections, or corpus-derived semantic axes derived via singular value decomposition (referred to here as EigAngle) \cite{deerwester1990lsa}. EigAngle derives angle parameters by projecting token-level statistics onto a low-dimensional semantic subspace obtained via truncated singular value decomposition (SVD) of a term co-occurrence matrix, with the resulting projections mapped to angular values through a monotonic normalization.

\paragraph{EigAngle: semantic angle construction.}
When semantic axes are available, angle parameters are derived from a corpus-dependent semantic representation obtained via singular value decomposition.

Let $\mathbf{E} \in \mathbb{R}^{V \times d_{\max}}$ denote the matrix of semantic axes, where each row corresponds to a token in the vocabulary. Given a window of tokens $\mathcal{W} = \{t_1, \dots, t_k\}$, we first compute the average semantic vector:
\[
\mathbf{v} = \frac{1}{k} \sum_{t_i \in \mathcal{W}} \mathbf{E}_{t_i}.
\]

This vector is then normalized via a per-dimension $z$-score transformation:
\[
\mathbf{z} = \frac{\mathbf{v} - \boldsymbol{\mu}}{\boldsymbol{\sigma} + \epsilon},
\]
where $\boldsymbol{\mu}$ and $\boldsymbol{\sigma}$ are the mean and standard deviation of the semantic axes, and $\epsilon$ is a small constant for numerical stability.

The resulting vector is truncated to the first $d$ components (with $d$ equal to the number of qubits; in the experiments reported in this work, $d=12$), and mapped to angular parameters:
\[
\boldsymbol{\theta} = \mathrm{clip}(\gamma \, \mathbf{z}_{1:d}, -\pi, \pi),
\]
where $\gamma$ is a scaling factor that controls the angular range.

If no semantic axes are available or no tokens in the window are covered by the vocabulary, a deterministic lexical fallback is used, ensuring that angle construction remains well-defined for all inputs.

Given an angle vector $\boldsymbol{\theta}$ derived from these sources, the encoder applies a parameterized feature mapping to produce a fixed-size representation. This mapping can be implemented either via a quantum circuit (Aer backend) or a classical surrogate. Formally, the encoder computes a feature vector $\mathbf{f}(\boldsymbol{\theta}) \in \mathbb{R}^F$ for each window. 

\paragraph{Quantum-inspired feature mapping.}
Given an angle vector $\boldsymbol{\theta} \in \mathbb{R}^{d}$, the encoder applies a parameterized transformation inspired by quantum circuits. This transformation can be expressed abstractly as:
\[
|\psi(\boldsymbol{\theta})\rangle = U(\boldsymbol{\theta}) |0\rangle,
\]
where $U(\boldsymbol{\theta})$ is a parameterized operator defined by a sequence of rotations and entangling transformations.
This construction is conceptually related to quantum feature maps used in kernel methods \cite{havlicek2019supervised}.

Feature extraction is performed by evaluating expectation values of a fixed set of observables:
\[
f_j(\boldsymbol{\theta}) = \langle \psi(\boldsymbol{\theta}) | O_j | \psi(\boldsymbol{\theta}) \rangle,
\]
where $\{O_j\}_{j=1}^{F}$ is a predefined collection of Pauli operators.

In the current implementation, the observable set includes single-qubit operators ($X_i, Y_i, Z_i$) and two-qubit interactions (e.g., $XX$, $YY$, $ZZ$) distributed deterministically across qubit pairs, producing a fixed number of features per window.

The transformation $U(\boldsymbol{\theta})$ is implemented using a hardware-efficient ansatz \cite{kandala2017hardware}, consistent with quantum-enhanced feature mappings \cite{havlicek2019supervised}, consisting of repeated layers of single-qubit rotations (e.g., $R_y$, $R_z$) followed by entangling gates arranged in a ring topology.

\begin{figure}[tbp]
\centering
\includegraphics[width=0.65
\linewidth]{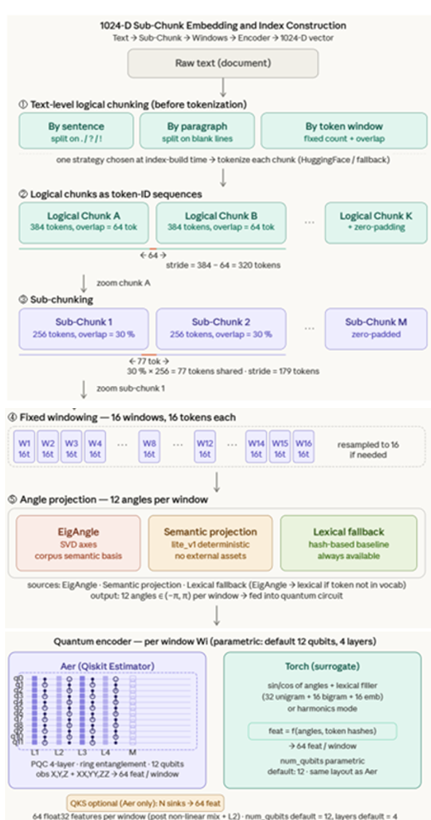}
\caption{QEMB encoding pipeline from raw text to 1024-dimensional sub-chunk embeddings. Documents are first segmented into logical chunks and sub-chunks. Each sub-chunk is processed through fixed windows whose features are obtained via angle projection and a quantum-inspired encoder implemented either with the Aer backend or a Torch surrogate.} 
\label{fig:qemb_pipeline}
\end{figure}

Figure~\ref{fig:qemb_pipeline} provides a schematic view of the QEMB encoding pipeline. The numerical values shown in the diagram illustrate one possible parameterization of the pipeline, while the framework itself supports different configurations depending on the experimental setup.

\paragraph{Text Segmentation Levels.}
The pipeline distinguishes two levels of textual segmentation. 
\emph{Logical chunks} are document-level segments produced by the corpus ingestion pipeline according to a configurable chunking policy and represent the primary units used for indexing and retrieval. 
Each logical chunk is further divided into smaller encoding units, referred to as \emph{sub-chunks}, which constitute the effective inputs processed by the QEMB encoder. 
Sub-chunk size and overlap are configurable parameters that control the granularity of the embedding process. 
This two-level segmentation allows the system to decouple document-level retrieval structure from the finer-grained encoding required for stable embedding construction.

\begin{figure}[!htbp]
\centering

\includegraphics[width=0.70\linewidth]{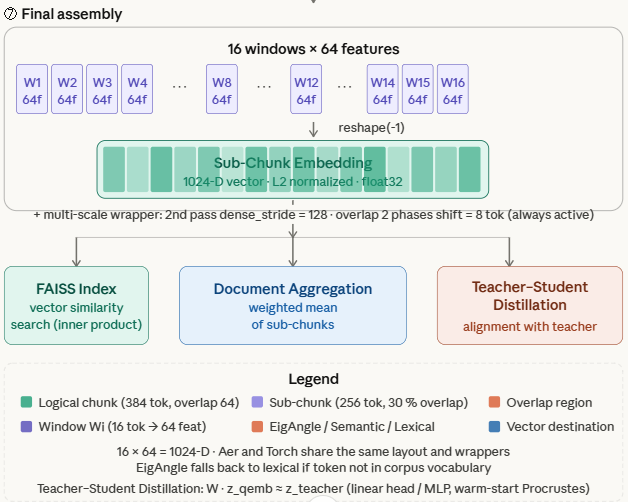}

\vspace{8mm}

\includegraphics[width=0.78\linewidth]{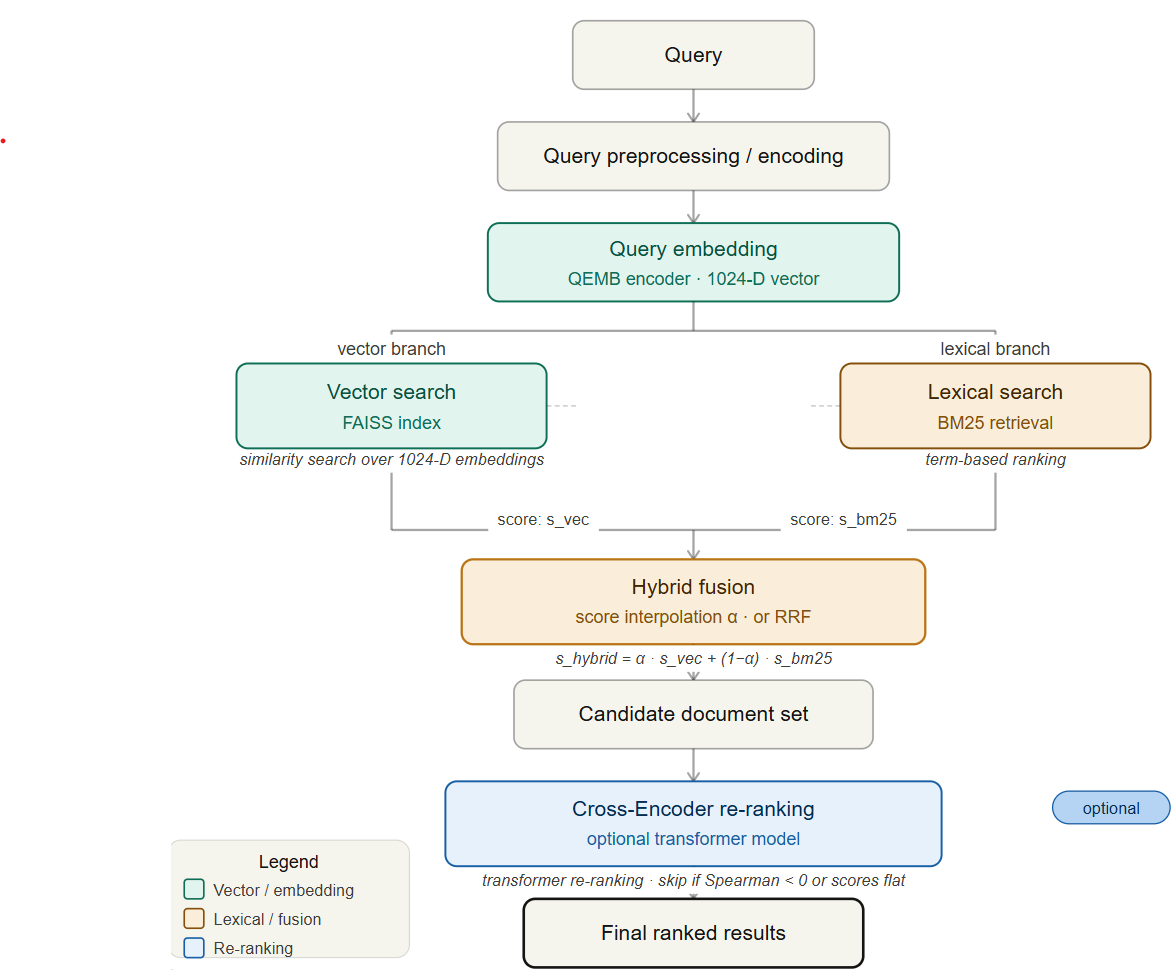}

\caption{Embedding construction and hybrid retrieval pipeline.
Top: embedding assembly.
Bottom: hybrid retrieval pipeline.
The upper diagram shows the assembly of the final 1024-dimensional sub-chunk embedding. The lower diagram illustrates the hybrid retrieval architecture combining vector similarity search over QEMB embeddings with lexical BM25 retrieval and optional cross-encoder re-ranking.}

\label{fig:qemb_embedding_and_retrieval}

\end{figure}

Each sub-chunk is partitioned into a fixed number of windows. For every window, the encoder produces a set of features derived either from a quantum-inspired circuit executed through the Aer backend or from a Torch-based surrogate implementation. The concatenation of window-level features produces a fixed-length embedding vector (1024 dimensions in the configuration used in this work).
The construction of the final embedding vector is summarized in the upper diagram of Figure~\ref{fig:qemb_embedding_and_retrieval}.

To improve representation stability, the pipeline applies overlap-based encoding and a dense multi-scale pass that aggregates multiple shifted views of the same text segment. The resulting vectors are aggregated and L2-normalized prior to indexing.

The quantum-inspired encoding components are implemented using Qiskit, with simulations executed on the Aer backend~\cite{qiskit2024}.

Where possible, GPU acceleration and parallel execution across CPU cores are leveraged to mitigate the computational cost of quantum circuit simulation.

These implementation choices are particularly relevant for components such as QKS and kernel-based transformations, where the number of circuit evaluations can become a dominant factor in runtime.

When the Aer backend is used, the pipeline can optionally incorporate random feature expansions inspired by Quantum Kitchen Sinks (QKS) \cite{wilson2018qks}. These transformations are applied at the window level to increase the diversity of feature projections.

The system also supports optional teacher--student distillation to align the embedding space with a stronger semantic model.

The resulting embeddings are used within a hybrid retrieval framework combining vector similarity with BM25 lexical scoring. Hybrid retrieval further includes candidate-union strategies and fusion mechanisms.
\paragraph{Similarity and score fusion.}
Given a query embedding $\mathbf{q}$ and a document (or sub-chunk) embedding $\mathbf{d}$, similarity is computed using inner product. Since all embeddings are L2-normalized, this corresponds to cosine similarity:
\[
s_{\mathrm{embed}}(q,d) = \langle \mathbf{q}, \mathbf{d} \rangle.
\]

Let $s_{\mathrm{BM25}}(q,d)$ denote the lexical score. Hybrid retrieval combines the two signals via score interpolation:
\[
s(q,d) = \alpha \, s_{\mathrm{embed}}(q,d) + (1-\alpha)\, s_{\mathrm{BM25}}(q,d),
\]
where $\alpha \in [0,1]$ controls the relative contribution of embedding-based and lexical signals.

In this work, $\alpha$ can be either fixed or dynamically adjusted based on query-dependent confidence signals, while maintaining the same interpolation form.

\paragraph{Candidate selection, fusion, and re-ranking.}
Retrieval is performed in two stages. First, a candidate set is constructed by combining top-$k$ results from embedding-based retrieval and BM25. This candidate-union strategy ensures that relevant documents are not discarded before score fusion.

Second, candidates are ranked using the hybrid score $s(q,d)$ defined above, where the balance between lexical and embedding signals is controlled by $\alpha$, which can be either static or dynamically gated. Rank-based fusion using Reciprocal Rank Fusion (RRF) is also considered as an alternative combination strategy.

Finally, an optional cross-encoder re-ranking stage based on transformer models \cite{reimers2019sentencebert} can be applied to the top-$k$ candidates. To avoid ranking instability, this stage is protected by safeguard conditions, which disable re-ranking when score distributions are degenerate (e.g., near-constant scores) or negatively correlated with the base ranking (Spearman correlation $< 0$).
\noindent The effectiveness of these components is evaluated in the experimental results reported below.

The overall hybrid retrieval architecture is illustrated in the lower diagram of Figure~\ref{fig:qemb_embedding_and_retrieval}.

A closely related implementation context is provided by Xu \textit{et al.}, 
who employ parameterized quantum circuits with angle encoding within a quantum recurrent architecture for text classification~\cite{xu2024recurrent}.

\subsubsection{Distillation}

To improve the semantic alignment of quantum-inspired embeddings, we consider a teacher--student distillation approach, where a classical embedding model (teacher) guides the construction of a transformed embedding space (student).

Let $\mathbf{e} \in \mathbb{R}^{1024}$ denote the original QEMB embedding and let $\mathbf{t} \in \mathbb{R}^{1024}$ denote the corresponding teacher embedding. Distillation applies a parametric mapping
\[
\mathbf{z} = g(\mathbf{e}),
\]
where $g$ is implemented either as a linear projection or as a Multi-Layer Perceptron (MLP), depending on the experimental configuration.

The distilled representation $\mathbf{z}$ is trained to approximate the teacher geometry through an alignment objective of the form
\[
\mathcal{L}_{\mathrm{align}} = \|\mathbf{z} - \mathbf{t}\|_2^2.
\]

Conceptually, the goal of distillation is not to replace the original quantum-inspired representation, but to improve its semantic organization while preserving the fixed-dimensional structure required by the retrieval pipeline.

In practice, however, the contribution of distillation remains modest. While some improvements in alignment can be observed, the overall behavior is strongly dependent on the dataset and query distribution. In particular, distillation does not fully compensate for the geometric limitations identified in Section~\ref{sec:direct_qemb}, and its effectiveness varies across domains and experimental conditions.

The impact of distillation on retrieval performance is further analyzed in the retrieval experiments reported below.

\FloatBarrier
\subsubsection{Dual-Channel Retrieval (Experimental)}

In addition to the standard embedding pipeline, we implemented an experimental dual-channel retrieval mode. In this configuration, two alternative embedding representations are computed: the standard QEMB representation (referred to as the \emph{current} channel) and an alternative amplitude-based representation (referred to as the \emph{amp} channel).

These representations can be combined at retrieval time using score aggregation strategies such as dual scoring or reciprocal rank fusion. The dual-channel configuration is primarily intended as an exploratory diagnostic tool rather than a stable component of the pipeline.

In the current implementation, the amplitude channel does not support the full set of pipeline features (e.g., distillation and certain retrieval modes) and is therefore evaluated only in controlled benchmark experiments.

\subsubsection{Why Quantum Kernels Are Not Included}

Quantum kernel methods are not integrated into the main pipeline because:

\begin{itemize}
\item they do not produce explicit embeddings;
\item they are not directly indexable;
\item their computational cost is prohibitive for large-scale retrieval.
\end{itemize}

They are therefore treated as exploratory tools rather than components of the production pipeline.

\subsection{Experimental Setup}

We evaluate retrieval on three controlled corpora designed to cover different semantic regimes and languages.

\begin{itemize}
\item \textbf{Technical corpus (Italian)}: 10 documents with technical content, on average one document is about 11450 tokens. This corpus is characterized by structured language and relatively low semantic ambiguity. 
Two query sets are considered: an initial set of 10 descriptive queries, and a larger set of 60 mixed queries including question, descriptive, and keyword formulations.

\item \textbf{Narrative corpus (English)}: 10 documents of narrative text, on average one document is about 8850 tokens. This corpus exhibits higher semantic variability and richer contextual dependencies.

\item \textbf{Legal corpus (Italian)}: 10 documents of legal text (judicial decisions), on average one document is about 4100 tokens. It is characterized by formal language, domain-specific terminology, and long-range dependencies.
\end{itemize}

The query sets include a mix of question, descriptive, and keyword formulations, with the exception of the initial technical experiment, which uses only descriptive queries. 
Short or keyword-style queries generally provide weaker and more ambiguous signals, leading to reduced ranking quality even for strong dense embeddings.

Although limited in size, these corpora enable controlled and interpretable evaluation of retrieval behavior across different domains and languages, allowing us to isolate the impact of embedding geometry on retrieval performance. The relatively small scale is also motivated by the computational cost of quantum circuit simulation (Aer backend), which makes large-scale experiments impractical in this setting. The number of queries varies across corpora, reflecting differences in domain complexity and query generation constraints. Since the goal is diagnostic analysis rather than standardized benchmarking, we do not enforce uniform query counts across datasets.

Unless otherwise stated, the retrieval results reported in the following sections refer to \textbf{document-level evaluation}. 
That is, relevance is assessed at the document level, and the reported metrics measure whether the correct document is retrieved and properly ranked. 
This choice reflects the main retrieval objective of the framework and keeps the comparison across corpora compact and interpretable. Furthermore, it reflects common practice in RAG systems, where retrieval is typically performed in two stages: first at the document level to identify relevant sources, and subsequently at a finer granularity (e.g., sub-chunks or passages) within the retrieved documents.

As a result, sub-chunk-level retrieval is not intended as a standalone retrieval task, but rather as a diagnostic probe of the embedding space. The observed degradation at this level therefore highlights structural limitations of the representation, rather than a mismatch with practical retrieval pipelines.

Sub-chunk-level retrieval is considered only as a \emph{diagnostic analysis}. 
Rather than reporting a full set of chunk-level metrics for all corpora, we provide selected examples to illustrate the collapse in ranking quality at finer granularity, which would otherwise make the experimental section unnecessarily heavy.

We evaluate the following document-level retrieval configurations: BM25, teacher embeddings (FAISS), QEMB embeddings (raw and distilled), and hybrid fusion with fixed interpolation parameter $\alpha$. The same value of $\alpha$ is used across all corpora for consistency, and is selected empirically as a representative trade-off rather than tuned per dataset.

BM25 is used as the primary lexical baseline, as it is widely recognized as a strong and well-established standard for document retrieval. Hybrid configurations are obtained via score-level interpolation between BM25 and embedding scores using a fixed value of $\alpha$ selected empirically. We do not perform an extensive sweep over $\alpha$, as the goal of this study is diagnostic rather than hyperparameter optimization.

All reported retrieval results include cross-encoder re-ranking applied to the top-$k$ candidates.

Retrieval performance is evaluated using standard document-level metrics, including Hit@K ($K \in \{1,3,5,10\}$), MRR@10, nDCG@10, and MAP@10.

Since each query is associated with a single relevant document, MRR and MAP coincide in this setting.

Chunking and segmentation parameters are not optimized and may vary between corpora depending on the characteristics of the dataset. 
In all cases, they are kept fixed within each experiment to isolate the effect of the embedding representations on retrieval behavior, rather than to maximize retrieval performance through indexing choices.

\subsection{Results on the Technical Corpus}

We first report retrieval results on the Italian technical corpus, using a controlled set of 10 queries. All experiments are conducted with a fixed retrieval pipeline, including top-$k$ candidate selection ($k=10$) followed by cross-encoder re-ranking.

\paragraph{Configuration.}
QEMB embeddings are constructed using a 1024-dim.\ embedding window-based encoding with overlap and multi-scale aggregation. Each document is segmented into logical chunks of approximately 384 tokens with 64-token overlap, and further divided into sub-chunks of 256 tokens with stride 179. 

Each sub-chunk is processed through fixed windows, whose features are derived via SVD-based semantic projections (EigAngle). Quantum-inspired transformations are simulated using the Aer backend with 12 qubits, 4 circuit layers, and 2048 shots. Multi-scale aggregation is enabled through dense overlapping windows.

When distillation is applied, the \emph{intfloat/multilingual-e5-large} model is used as teacher, and a 1024-dimensional MLP head is trained to align the QEMB representation space. This choice provides a strong multilingual baseline, although alternative teacher models could be considered. The distilled embeddings achieve moderate alignment with the teacher space ($r = 0.46$, MAE = 0.08). Here, $r$ denotes the Pearson correlation coefficient.

\paragraph{Main results.}
Table~\ref{tab:technical_results} reports the main retrieval metrics.

\begin{table}[t]
\centering
\caption{Retrieval results on the technical corpus (10 descriptive queries). Hybrid configurations use score-level interpolation with $\alpha = 0.7$.}
\label{tab:technical_results}
\footnotesize
\setlength{\tabcolsep}{5pt}
\begin{tabular}{lccccccc}
\hline
\textbf{Method} & H@1 & H@3 & H@5 & H@10 & nDCG & MRR & MAP \\
\hline
Teacher & 0.90 & 1.00 & 1.00 & 1.00 & 0.963 & 0.95 & 0.95 \\
BM25 & 0.80 & 0.90 & 1.00 & 1.00 & 0.893 & 0.858 & 0.858 \\
QEMB (raw) & 0.30 & 0.30 & 0.50 & 1.00 & 0.542 & 0.41 & 0.41 \\
QEMB (dist.) & 0.00 & 0.40 & 0.40 & 1.00 & 0.409 & 0.24 & 0.24 \\
Hybrid (raw) & 0.90 & 1.00 & 1.00 & 1.00 & 0.963 & 0.95 & 0.95 \\
Hybrid (dist.) & 0.70 & 1.00 & 1.00 & 1.00 & 0.889 & 0.85 & 0.85 \\
\hline
\end{tabular}
\end{table}

\paragraph{Robustness check (60 queries).}
The query set includes a mix of question, descriptive, and keyword queries.

\begin{table}[t]
\centering
\caption{Retrieval results on the technical corpus (60 mixed queries). Hybrid configurations use score-level interpolation with $\alpha = 0.7$.}
\label{tab:technical_results_60}
\footnotesize
\setlength{\tabcolsep}{5pt}
\begin{tabular}{lccccccc}
\hline
\textbf{Method} & H@1 & H@3 & H@5 & H@10 & nDCG & MRR & MAP \\
\hline
Teacher & 0.62 & 0.75 & 0.88 & 1.00 & 0.787 & 0.721 & 0.721 \\
BM25 & 0.53 & 0.75 & 0.85 & 1.00 & 0.748 & 0.669 & 0.669 \\
QEMB (raw) & 0.08 & 0.35 & 0.55 & 1.00 & 0.455 & 0.292 & 0.292 \\
QEMB (dist.) & 0.20 & 0.38 & 0.63 & 1.00 & 0.516 & 0.372 & 0.372 \\
Hybrid (raw) & 0.63 & 0.75 & 0.85 & 1.00 & 0.795 & 0.732 & 0.732 \\
Hybrid (dist.) & 0.50 & 0.67 & 0.78 & 1.00 & 0.714 & 0.627 & 0.627 \\
\hline
\end{tabular}
\end{table}

\paragraph{Analysis.}
The results in Table~\ref{tab:technical_results} highlight a clear gap between lexical and quantum-inspired representations. BM25 provides a strong baseline, achieving high recall within the top-5 results and competitive ranking quality, which is expected in structured technical domains with low lexical ambiguity.

In contrast, QEMB embeddings exhibit significantly weaker performance when used in isolation. Although relevant documents are consistently retrieved within the top-10 (Hit@10 = 1.00), the ranking quality is poor, as reflected by low Hit@1 and nDCG values. This indicates that QEMB embeddings retain weak relevance signals but fail to provide a reliable ordering of candidates.

Distillation does not consistently improve this behavior. While minor variations across ranks are observed (e.g., slight improvements at intermediate cutoffs), the overall ranking quality remains unstable and, in some cases, degrades further. This suggests that simple projection-based alignment is insufficient to recover a robust semantic structure in the embedding space.

A different picture emerges in the hybrid setting. When combined with BM25 and re-ranked by a cross-encoder, raw QEMB embeddings contribute to a retrieval signal that matches the teacher baseline. This indicates that, despite their poor standalone performance, QEMB embeddings encode complementary information that can be exploited in a hybrid pipeline.

However, this effect is not preserved under distillation. Hybrid retrieval with distilled embeddings performs worse than both the raw hybrid configuration and the BM25 baseline, suggesting that distillation may distort the weak but useful signals present in the original QEMB representation.

Taken together, these results indicate that quantum-inspired embeddings, in their current form, are not sufficient as standalone retrieval representations, but can act as auxiliary signals in hybrid retrieval under specific conditions.

The larger 60-query experiment reported in Table~\ref{tab:technical_results_60} broadly confirms this picture, while also refining it in important ways. In particular, it shows that the relative behavior of BM25, teacher embeddings, raw QEMB, and distilled QEMB depends on query formulation more than the initial 10-query setting suggested. The general weakness of standalone QEMB is confirmed, but the role of distillation and hybrid fusion turns out to be more nuanced than in the smaller descriptive-only experiment.

\paragraph{Analysis (60-query robustness setting).}
The results in Table~\ref{tab:technical_results_60} provide a more robust view of retrieval behavior on the technical corpus than the initial 10-query experiment reported in Table~\ref{tab:technical_results}. The larger query set is more heterogeneous, including question, descriptive, and keyword formulations, whereas the initial experiment used only descriptive queries.

A first difference concerns the relative behavior of lexical and dense retrieval. In the 10-query setting, BM25 appeared particularly strong, which is consistent with the descriptive nature of the queries and the structured terminology of the corpus. In the 60-query setting, teacher embeddings become comparatively stronger, while BM25 remains competitive but no longer dominates to the same extent. This suggests that the larger mixed query set reduces the advantage of purely lexical matching and makes semantic retrieval more relevant.

QEMB embeddings remain clearly weaker than both BM25 and teacher embeddings. In contrast to the initial 10-query experiment, where the descriptive formulation of the queries allowed QEMB to retain a limited but non-negligible ranking signal, the 60-query mixed setting exposes a more pronounced degradation, especially for the raw embeddings. This indicates that QEMB is more sensitive to heterogeneous query formulations and struggles to provide stable ranking signals when lexical overlap becomes less informative or when the query is short and ambiguous.

Distillation improves standalone QEMB retrieval in the 60-query setting, yielding better ranking quality than the raw representation. This is an important difference with respect to the smaller experiment, where the gains from distillation were weaker and less convincing. However, even with this improvement, distilled QEMB remains substantially below both teacher and BM25 retrieval, confirming that the semantic structure recovered by distillation is still insufficient for strong standalone performance.

In the hybrid setting, the results are also more nuanced than in the 10-query experiment. Hybrid retrieval with raw QEMB reaches performance comparable to the teacher baseline and slightly above it on some metrics, suggesting that raw quantum-inspired embeddings can still provide a useful complementary signal when combined with BM25 over a more varied query distribution. By contrast, the distilled hybrid configuration performs noticeably worse than the raw hybrid one, even though distilled QEMB is stronger than raw QEMB in isolation. This indicates that improving the standalone embedding space through distillation does not necessarily preserve the weak but useful complementary information exploited by hybrid fusion.

Collectively, the 60-query experiment refines the interpretation of the technical corpus results. It confirms the general weakness of QEMB as a standalone retrieval representation, but also shows that query formulation matters substantially: descriptive queries tend to favor lexical retrieval, while a more heterogeneous query set reveals a stronger role for teacher embeddings and a more selective, less stable contribution from quantum-inspired representations.

\subsection{Results on the Narrative Corpus}

We report retrieval results on the English narrative corpus composed of 10 documents. This corpus exhibits higher semantic variability and contextual richness compared to the technical setting, making lexical matching less effective and placing greater emphasis on semantic representations.

\paragraph{Configuration.}
We adopt the same evaluation protocol and overall retrieval pipeline used in the previous experiments, including top-$k$ candidate selection ($k=10$) followed by cross-encoder re-ranking. 
Evaluation is performed at the \textbf{document level}, consistently with the technical corpus.

QEMB embeddings are constructed using a 1024-dim.\ embedding window-based encoding with overlap and multi-scale aggregation. 
Each document is segmented into logical chunks of approximately 128 tokens with 32-token overlap, and further divided into sub-chunks of 128 tokens with stride 96.

Each sub-chunk is processed through fixed windows, whose features are derived via semantic projections (EigAngle). 
Quantum-inspired transformations are simulated using the Aer backend with 12 qubits, 4 circuit layers, and analytic evaluation (no sampling noise). 
Multi-scale aggregation is enabled through dense overlapping windows (stride 128), combined with a two-phase overlap strategy with an 8-token shift.

When distillation is applied, the \emph{intfloat/multilingual-e5-large} model is used as teacher, and a 1024-dimensional MLP head is trained to align the QEMB representation space.

The final distilled embeddings achieve weak alignment with the teacher space ($r = 0.18$, MAE = 0.10).

\paragraph{Main results.}
Table~\ref{tab:narrative_results} reports the main retrieval metrics.

\begin{table}[t]
\centering
\caption{Retrieval results on the narrative corpus (60 mixed queries). All methods use top-$k=10$ retrieval followed by cross-encoder re-ranking.}
\label{tab:narrative_results}
\footnotesize
\setlength{\tabcolsep}{5pt}
\begin{tabular}{lccccccc}
\hline
\textbf{Method} & H@1 & H@3 & H@5 & H@10 & nDCG & MRR & MAP \\
\hline
Teacher & 0.37 & 0.58 & 0.75 & 1.00 & 0.650 & 0.542 & 0.542 \\
BM25 & 0.50 & 0.65 & 0.78 & 1.00 & 0.714 & 0.627 & 0.627 \\
QEMB (raw) & 0.12 & 0.35 & 0.58 & 1.00 & 0.480 & 0.324 & 0.324 \\
QEMB (dist.) & 0.20 & 0.33 & 0.45 & 1.00 & 0.502 & 0.358 & 0.358 \\
Hybrid (raw) & 0.42 & 0.62 & 0.75 & 1.00 & 0.674 & 0.574 & 0.574 \\
Hybrid (dist.) & 0.43 & 0.60 & 0.72 & 1.00 & 0.672 & 0.573 & 0.573 \\
\hline
\end{tabular}
\end{table}

\paragraph{Analysis.}
The results on the narrative corpus highlight a more challenging retrieval setting compared to the technical and legal domains. 

First, BM25 remains a strong baseline, achieving higher ranking quality than the teacher embeddings. However, its performance degrades compared to the technical corpus, reflecting the higher semantic variability and reduced lexical alignment typical of narrative text.

Teacher embeddings also exhibit weaker ranking performance than in previous settings. Although relevant documents are consistently retrieved within the top-10 (Hit@10 = 1.00), the lower Hit@1 and MRR values indicate increased difficulty in correctly ordering candidates, suggesting that even strong dense representations struggle in this domain.

QEMB embeddings show a clear degradation when used in isolation. The raw configuration yields poor ranking quality, with a significant drop across all metrics. Distillation provides only a limited improvement, confirming that alignment with the teacher space does not translate into robust retrieval performance.

In the hybrid setting, combining BM25 with QEMB embeddings improves performance compared to standalone QEMB. However, hybrid configurations remain below the BM25 baseline, indicating that the contribution of QEMB embeddings is weaker in this domain. The difference between raw and distilled hybrid configurations is minimal, suggesting that distillation does not provide a consistent advantage.

Overall, the narrative corpus confirms the trends observed in other settings while amplifying their effects: dense and quantum-inspired embeddings struggle to provide stable ranking signals, and hybrid retrieval does not fully compensate for these limitations.

\subsection{Results on the Legal Corpus}

We evaluate retrieval on the Italian legal corpus composed of 10 judicial documents. 
This domain is characterized by formal language, domain-specific terminology, and long-range dependencies, providing a realistic testbed for structured document retrieval.

\paragraph{Configuration.}
We adopt the same evaluation protocol and overall retrieval pipeline used in the previous experiments, including top-$k$ candidate selection ($k=10$) followed by cross-encoder re-ranking. 
Evaluation is performed at the \textbf{document level}, consistently with the technical corpus.

The legal corpus is evaluated using a larger set of 84 mixed  queries, providing a more statistically robust assessment of retrieval performance compared to the smaller setups.

QEMB embeddings are constructed using a 1024-dim.\ embedding window-based encoding with overlap and multi-scale aggregation. 
Each document is segmented into logical chunks of approximately 384 tokens with 64-token overlap, and further divided into sub-chunks of 256 tokens with stride 179.

Each sub-chunk is processed through fixed windows, whose features are derived via semantic projections (EigAngle). 
Quantum-inspired transformations are simulated using the Aer backend with 12 qubits, 4 circuit layers, and analytic evaluation (no sampling noise). 
Multi-scale aggregation is enabled through dense overlapping windows (stride 128), combined with a two-phase overlap strategy with an 8-token shift.

When distillation is applied, the \emph{intfloat/multilingual-e5-large} model is used as teacher, and a 1024-dimensional MLP head is trained to align the QEMB representation space.

The final distilled embeddings achieve strong alignment with the teacher space ($r = 0.88$, MAE = 0.03), although, as shown in the results, this geometric alignment does not necessarily translate into improved retrieval performance.

\begin{table}[t]
\centering
\caption{Retrieval results on the legal corpus (84 mixed queries). Hybrid configurations use score-level interpolation with $\alpha = 0.7$.}
\label{tab:legal_results}
\footnotesize
\setlength{\tabcolsep}{5pt}
\begin{tabular}{lccccccc}
\hline
\textbf{Method} & H@1 & H@3 & H@5 & H@10 & nDCG & MRR & MAP \\
\hline
Teacher & 0.96 & 0.99 & 1.00 & 1.00 & 0.984 & 0.979 & 0.979 \\
BM25 & 1.00 & 1.00 & 1.00 & 1.00 & 1.000 & 1.000 & 1.000 \\
QEMB (raw) & 0.11 & 0.37 & 0.61 & 1.00 & 0.474 & 0.315 & 0.315 \\
QEMB (dist.) & 0.19 & 0.42 & 0.67 & 1.00 & 0.524 & 0.380 & 0.380 \\
Hybrid (raw) & 0.99 & 1.00 & 1.00 & 1.00 & 0.996 & 0.994 & 0.994 \\
Hybrid (dist.) & 0.92 & 0.99 & 0.99 & 1.00 & 0.966 & 0.954 & 0.954 \\
\hline
\end{tabular}
\end{table}

\paragraph{Analysis.}
The results reported in Table~\ref{tab:legal_results} highlight a clear and consistent pattern across all configurations.

First, BM25 provides an exceptionally strong baseline in the legal domain, achieving perfect performance across all metrics. This behavior is expected given the highly structured nature of legal language, where terminology is precise and lexical overlap between queries and relevant documents is high. In this setting, lexical matching alone is sufficient to retrieve and correctly rank the target documents.

Teacher embeddings also achieve near-perfect performance, confirming that the retrieval task is well-defined and semantically coherent. However, their performance remains slightly below BM25, suggesting that dense retrieval may introduce minor deviations in candidate ranking compared to exact lexical matching.

In contrast, QEMB embeddings exhibit significantly weaker performance when used in isolation. Although relevant documents are consistently retrieved within the top-10 results (Hit@10 = 1.00), ranking quality is poor, as reflected by low Hit@1 and MRR values. This behavior is consistent with the findings observed in the technical corpus and indicates that QEMB embeddings retain weak relevance signals but fail to provide a reliable ordering of candidates.

Distillation improves the standalone performance of QEMB, increasing all ranking metrics. However, the overall performance remains far below both BM25 and teacher embeddings. This confirms that, despite strong geometric alignment with the teacher space, the distilled representations do not preserve the local neighborhood structure required for effective retrieval.

A markedly different behavior emerges in the hybrid setting. When combined with BM25 using score-level interpolation, QEMB embeddings contribute to a retrieval signal that achieves near-perfect performance. In particular, the hybrid configuration with raw embeddings closely matches BM25 and slightly improves over the teacher baseline in ranking metrics, indicating that QEMB can provide complementary information when integrated with a strong lexical signal.

However, this effect is reduced when using distilled embeddings. While hybrid (distilled) still achieves high performance, it is consistently worse than the raw hybrid configuration. This suggests that distillation, while improving global alignment, may distort the aspects of the representation that are beneficial for hybrid fusion.

Overall, the results confirm and strengthen the observations from the technical corpus. QEMB embeddings are not sufficient as standalone retrieval representations, due to limited ranking stability and weak discriminative power. However, they can act as auxiliary signals in hybrid retrieval, provided that their original structure is preserved. Furthermore, the results highlight a key limitation of distillation: improvements in geometric alignment do not necessarily translate into better retrieval performance and may even degrade hybrid effectiveness.

\subsection{Sub-chunk-Level Diagnostic Analysis}

To further investigate the behavior of the embedding representations, we conduct a diagnostic analysis at the sub-chunk level. 
Unlike document-level retrieval, this setting evaluates the ability of the model to correctly identify the specific sub-chunk containing the relevant information.

The experiment is performed on the technical corpus using 30 queries at the sub-chunk level. 
The same retrieval configuration adopted in the document-level evaluation is used here, including top-$k$ retrieval and cross-encoder re-ranking.

\paragraph{Results.}
Table~\ref{tab:subchunk_results} reports the retrieval performance for teacher embeddings and QEMB (raw) representations.

\begin{table}[t]
\centering
\caption{Sub-chunk-level retrieval results on the technical corpus (30 queries).}
\label{tab:subchunk_results}
\footnotesize
\setlength{\tabcolsep}{5pt}
\begin{tabular}{lccccccc}
\hline
\textbf{Method} & H@1 & H@3 & H@5 & H@10 & nDCG & MRR & MAP \\
\hline
Teacher & 0.03 & 0.30 & 0.43 & 0.67 & 0.251 & 0.206 & 0.192 \\
QEMB (raw) & 0.00 & 0.00 & 0.03 & 0.07 & 0.026 & 0.014 & 0.014 \\
\hline
\end{tabular}
\end{table}

\paragraph{Analysis.}
The results highlight a substantial increase in difficulty when moving from document-level to sub-chunk-level retrieval. 
Even the teacher embeddings exhibit a marked degradation in ranking quality, with low Hit@1 and moderate recall at higher cutoffs. 
This reflects the intrinsic challenge of fine-grained retrieval, where the task requires precise localization rather than coarse document matching.

In contrast, QEMB embeddings exhibit a near-complete collapse in performance. 
Relevant sub-chunks are rarely retrieved within the top ranks, and the overall ranking quality is extremely low across all metrics. 
This behavior is consistent with the previously observed limitations of the embedding space, including weak discriminative power and unstable similarity structure.

Overall, this diagnostic analysis confirms that the limitations of QEMB are amplified at finer granularity. 
While document-level retrieval may partially mask these issues, sub-chunk-level evaluation exposes a fundamental inability to capture and preserve local semantic relevance.

\section{Discussion and Limitations}
\label{sec:limitations}

The experimental results presented across technical, narrative, and legal corpora consistently highlight a set of structural limitations of quantum-inspired embeddings in retrieval tasks.

A first observation concerns the mismatch between geometric alignment and retrieval effectiveness. Although distillation improves alignment with the teacher space in some settings, this does not reliably translate into better ranking performance. From a broader perspective, these limitations are consistent with known challenges in parameterized quantum models. In particular, variational quantum circuits are known to suffer from barren plateau phenomena, where gradients vanish and optimization becomes ineffective \cite{mcclean2018barren}. While the present work does not involve training quantum circuits, the observed lack of discriminative structure and similarity compression suggests that related expressivity limitations may arise even in quantum-inspired feature mappings.

This suggests that distillation acts as a geometric projection that improves global similarity alignment while potentially distorting local neighborhood structure. As a result, retrieval performance may degrade despite improved correlation with the teacher space, highlighting a fundamental mismatch between global alignment objectives and ranking-sensitive geometry.

A second recurring phenomenon is the instability of ranking signals. Quantum-inspired embeddings often exhibit distance compression effects, producing weak or compressed similarity distributions, which result in poor discrimination between relevant and non-relevant documents. This effect is further reflected in the behavior of the cross-encoder re-ranking stage, which is frequently disabled due to negative or inconsistent rank correlation with the initial candidate ordering. This indicates that the underlying representation does not provide a stable basis for subsequent refinement.

Third, hybrid retrieval generally improves performance over standalone QEMB embeddings, confirming that lexical signals remain dominant and reliable. However, the contribution of quantum-inspired embeddings in hybrid configurations is limited and dataset-dependent, and does not systematically lead to improvements over strong lexical baselines.

These observations recur across all corpora, despite differences in language, domain, and query formulation. 
Absolute performance varies depending on the characteristics of the data set, but qualitative trends remain broadly similar, suggesting that the observed behavior reflects intrinsic properties of the embedding construction rather than purely corpus-specific artifacts. 
Although experiments are conducted on controlled datasets, the goal of this work is not to provide exhaustive benchmarking but to identify structural behaviors that remain consistent across different settings. 
This evidence is consistent with recent work showing that conclusions on quantum and quantum-inspired models can be strongly influenced by benchmarking choices, particularly in terms of baseline selection and experimental setup \cite{bowles2024better}.

From a modeling perspective, a key limitation lies in the static nature of the encoding. The mapping from text to embedding is deterministic and does not incorporate contextual adaptation or learned semantic transformations at the representation level. As a result, the model struggles to capture fine-grained semantic distinctions and to maintain a meaningful similarity structure across diverse inputs.

Finally, it is important to note that the experimental settings are designed as controlled diagnostic case studies rather than optimized retrieval pipelines. Chunking strategies, indexing parameters, and model configurations are intentionally not tuned to maximize performance. The goal is to isolate the behavior of the representations under consistent conditions, enabling a clearer interpretation of their strengths and limitations.

\section{Conclusions and Future Work}
\label{sec:conclusions}

This work presented a controlled experimental framework for evaluating quantum-inspired 1024-dimensional embeddings in document retrieval settings. 
Instead of aiming for state-of-the-art performance, the study focused on understanding the representational properties of these embeddings and their impact on retrieval behavior across different domains and query formulations.

The experimental results consistently show that quantum-inspired embeddings exhibit fundamental limitations. 
At the geometric level, the embedding space fails to preserve meaningful similarity structure, leading to weak alignment with reference semantic signals and to distance compression effects. 
A particularly critical phenomenon identified in this work is what we term a \emph{pathological inversion of similarity structure}, in which semantically related and unrelated inputs are no longer consistently ordered in the embedding space. 

Future research should therefore investigate encoding strategies that explicitly prevent such inversions, for example through geometric regularization, ranking-aware objectives, or constraints on similarity monotonicity, in order to preserve meaningful neighborhood structure.

These issues translate directly into retrieval performance, where QEMB embeddings provide unstable and poorly discriminative ranking signals.

Distillation improves global alignment with teacher embeddings in some configurations, but does not reliably translate into better retrieval performance. 
This highlights a gap between global similarity matching and the preservation of local neighborhood structure required for effective ranking.

Hybrid retrieval partially mitigates these limitations by combining lexical and embedding-based signals. 
In several settings, hybrid configurations achieve performance comparable to strong baselines, indicating that quantum-inspired embeddings can provide complementary signals. 
However, these improvements are not consistent across datasets, with BM25 often remaining the dominant contributor. 
This suggests that quantum-inspired embeddings, in their current form, act primarily as auxiliary signals rather than standalone retrieval representations.

A key insight of this work is that these limitations become more evident at finer granularity. 
While document-level retrieval may partially mask representation weaknesses, sub-chunk-level evaluation exposes a clear collapse in performance, indicating a fundamental inability to capture local semantic relevance.

From a broader perspective, the observed behavior is consistent across corpora and settings, suggesting that the limitations arise from intrinsic properties of the encoding rather than from dataset-specific factors or parameter choices. 

More broadly, these findings suggest that the observed degradation is not merely an empirical artifact, but reflects structural constraints in the construction of high-dimensional quantum-inspired embeddings that preserve retrieval-relevant similarity. In this sense, the limitations identified in this work help define the practical boundaries of quantum-inspired approaches in document retrieval, suggesting that their role is better suited to auxiliary or hybrid components rather than standalone embedding models.

\paragraph{Future Work.}
Future research directions can be grouped into three main areas.

First, improving the representational capacity of the embedding space. 
This may involve moving beyond static, deterministic mappings and introducing adaptive or learned transformations capable of preserving semantic structure more effectively.

Second, incorporating ranking-aware objectives during training. 
Current distillation strategies focus on global alignment, but do not explicitly optimize for retrieval metrics. 
Integrating contrastive or ranking-based losses could help bridge the gap between geometric similarity and retrieval effectiveness.

Third, exploring alternative integration strategies within hybrid pipelines. 
Rather than simple score-level fusion, more structured interactions between lexical and embedding signals could be investigated, including learned fusion mechanisms or multi-stage retrieval architectures.

Finally, quantum-inspired kernels and transformations remain an interesting direction for theoretical analysis, although their computational cost and lack of indexable representations currently limit their applicability in large-scale retrieval systems.

\section*{Acknowledgment}
We acknowledge financial support under the PR Puglia FESR FSE+ 2021-2027 - European Regional Development Fund, Net Service project QUICK SHIELD of Puglia Region, CUP:B85H24000920007.

\bibliographystyle{IEEEtran}
\bibliography{references_qemb}

\end{document}